\documentclass[twocolumn, iop]{openjournal}
\usepackage{xcolor}
\definecolor{xlinkcolor}{cmyk}{1,1,0,0}
\usepackage[bookmarks=true,         
    pdfnewwindow=true,      
   colorlinks=true,    
  linkcolor=xlinkcolor,     
 citecolor=xlinkcolor,     
filecolor=xlinkcolor,  
urlcolor=xlinkcolor,      
final=true,
 ]{hyperref}
\usepackage{amsmath}	
\usepackage{amssymb}	

\usepackage{physics}
\usepackage{orcidlink}

\newcommand{\be}{\begin{equation}}
\newcommand{\ee}{\end{equation}}
\newcommand{\units}[1]{\ensuremath{\, \mathrm{#1}}}  
\newcommand{\Msun}{{\rm M}_\odot}
\newcommand{\Mjup}{{\rm M}_{\rm Jup}}

\newcommand{\Rjup}{{\rm R}_{\rm Jup}}

\newcommand{\uvec}[1]{\boldsymbol{\hat{\textbf{#1}}}}
\newcommand{\cprod}[2]{\uvec{#1} \times \uvec{#2}}
\newcommand{\dprod}[2]{\uvec{#1} \cdot \uvec{#2}}
\newcommand{\sxl}{(\cprod{s}{l})}
\newcommand{\sdotl}{(\dprod{s}{l})}
\newcommand{\sxn}{(\cprod{s}{n})}
\newcommand{\sdotn}{(\dprod{s}{n})}
\newcommand{\sxk}{(\cprod{s}{k})}

\newcommand{\lxs}{(\cprod{l}{s})}
\newcommand{\lDots}{(\dprod{l}{s})} 
\newcommand{\lxn}{(\cprod{l}{n})}
\newcommand{\ldotn}{(\dprod{l}{n})}
\newcommand{\lxk}{(\cprod{l}{k})}

\usepackage[T1]{fontenc}

\newcommand{\orcidauthor}[3]{\author{\href{http://orcid.org/#1}{#2$^{#3}$}}}

\emergencystretch=\maxdimen
\begin{document}

\title{\vspace{-0.8cm} A potential exomoon from the predicted planet obliquity of $\beta$~Pictoris~b \vspace{-1.5cm}}

\orcidauthor{0000-0001-7739-9767}{Michael Poon}{1,*}
\orcidauthor{0000-0003-1927-731X}{Hanno Rein}{2,1}
\orcidauthor{0000-0002-0924-8403}{Dang Pham}{1}

\affiliation{$^{1}$Department of Astronomy and Astrophysics, University of Toronto, Toronto, Ontario, M5S 3H4, Canada}

\affiliation{$^{2}$Department of Physical and Environmental Sciences, University of Toronto Scarborough, Toronto, Ontario, M1C 1A4, Canada}

\thanks{$^*$Email: \href{mailto:michael.poon@astro.utoronto.ca}{michael.poon@astro.utoronto.ca}}

\begin{abstract}

Planet obliquity is the alignment or misalignment of a planet spin axis relative to its orbit normal.  
In a multiplanet system, this obliquity is a valuable signature of planet formation and evolutionary history.
The young $\beta$~Pictoris system hosts two coplanar super-Jupiters and upcoming JWST observations of this system will constrain the obliquity of the outer planet, $\beta$~Pictoris b. This will be the first planet obliquity measurement in an extrasolar, multiplanet system.
First, we show that this new planet obliquity is likely misaligned by using a wide range of simulated observations in combination with published measurements of the system. 
Motivated by current explanations for the tilted planet obliquities in the Solar System, we consider collisions and secular spin-orbit resonances.
While collisions are unlikely to occur, secular spin-orbit resonance modified by the presence of an exomoon around the outer planet can excite a large obliquity.
The largest induced obliquities ($\sim 60^\circ$) occur for moons with at least a Neptune-mass and a semimajor axis of $0.03-0.05 \units{au}$ ($40-70$ planet radii).
For certain orbital alignments, such a moon may observably transit the planet (transit depth of $3-7\%$, orbital period of $3-7$ weeks). Thus, a nonzero obliquity detection of $\beta$~Pictoris~b implies that it may host a large exomoon.
Although we focus on the $\beta$ Pictoris system, the idea that the presence of exomoons can excite high obliquities is very general and applicable to other exoplanetary systems.
\end{abstract}

\section{Introduction} \label{sec:intro}

The measurement of a planet obliquity, which is the alignment or misalignment between a planet spin axis and its orbit normal, provides a new avenue to explore an exoplanet's dynamics and evolutionary history. Although planet obliquities in our Solar System are precisely measured, constraints on \textit{exoplanet} obliquities are broad and limited. This is due to the challenge of requiring multiple observationally intensive measurements: high-resolution spectra and space-based photometry to constrain the planet spin axis, and high-precision astrometry over a near-decade baseline to constrain the orbital plane. To date, there are only four extrasolar systems with planet obliquity measurements (2M0122-2439~b, \citealt{Bryan+2020_Obliquity}; HD~106906~b, \citealt{Bryan+2021}; AB~Pictoris~b, \citealt{Palma-Bifani+2023}; and VHS~1256-1257~b, \citealt{Poon+2024}), each with a single confirmed planetary-mass companion ($M \sim 10\units{\Mjup}$) at wide separation ($\gtrsim 50\units{au}$).

The $\beta$~Pictoris system hosts two confirmed planets ($\beta$~Pictoris~b~and~c) on near coplanar orbits, residing at $10.26 \pm 0.10 \units{au}$ and $2.74 \pm 0.03 \units{au}$ from the host star, respectively \citep{Brandt+2021_OrbitFits}. $\beta$~Pictoris~b is of particular interest, because an accepted James Webb Space Telescope proposal (ID: 4758, Cycle 3) is anticipated to measure the rotation period of this planet, which is the last ingredient\footnote{
    Constraining a planet obliquity requires (1) an orbital inclination from astrometry, (2) a line-of-sight projected equatorial velocity from high resolution spectroscopy, and (3) a rotation period from time-resolved photometry (see Fig. 1 of \citealt{Poon+2024} for details).
}
required for a planet obliquity constraint of this planet. This would be the first extrasolar planet obliquity measurement in a multiplanet system, providing a unique opportunity to compare with giant planets in our Solar System. Complementary to this is the \textit{stellar} obliquity of $\beta$~Pictoris~b, where \citet{Kraus+2020} found that the angular momentum vectors of the stellar spin, outer planet orbit normal, and outer debris disk are well aligned with mutual inclinations $\leq 3^\circ \pm 5^\circ$.

The standard expectation is that giant planets forming through core accretion start with near-zero planet obliquity, since formation is slow and the current spin/orbital angular momentum should be similar to its natal circumstellar disk \citep[e.g.][]{Dones+Tremaine1993, Johansen+Lacerda2010}. Although the planets in our Solar System have nearly coplanar orbits and low stellar obliquities ($\lesssim 7^\circ$, \citealt{Beck+Giles2005}), their spin vectors often deviate from alignment to their orbit normal. For example, Saturn, Neptune and Uranus have observed planet obliquities of $27^\circ, 28^\circ$, and $98^\circ$, respectively. Explanations for these nonzero planet obliquities include collisions \citep{Kegerreis+2018}, and the onset of secular spin-orbit resonance induced either by moon migration \citep{Saillenfest+2021_Saturn} or planet migration \citep{Lu+Laughlin2022, Lu+2024}. 

\begin{figure}
    \centering
    \includegraphics[width=1.0\linewidth]{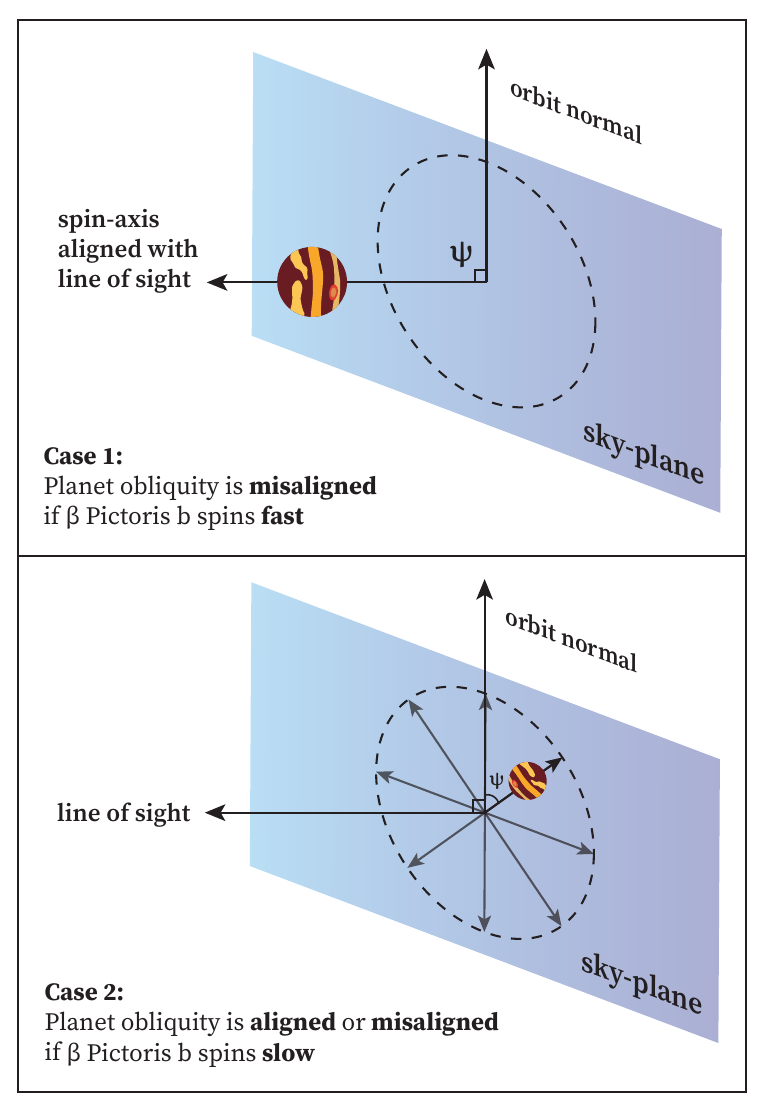}
    \caption{Schematic diagram of planet obliquity outcomes for $\beta$~Pictoris~b in the limit of an exactly pole-on (Case 1, top) or edge-on (Case 2, bottom) spin axis along our line of sight. In Case 1, if $\beta$~Pictoris~b spins \textit{fast} (i.e. $v \gg v\sin{i_p}$), the resulting spin axis inclination $i_p \sim 0^\circ$ implies a misaligned planet obliquity. In Case 2, if $\beta$~Pictoris~b spins \textit{slow} (i.e. $v \sim v\sin{i_p}$), the spin axis and orbit normal both lie in the sky-plane, and therefore an aligned or misaligned planet obliquity is possible.
    }
    \label{fig:skyplane_obliquity}
\end{figure}

\begin{figure*}
    \centering
    \includegraphics[width=1.0\linewidth]{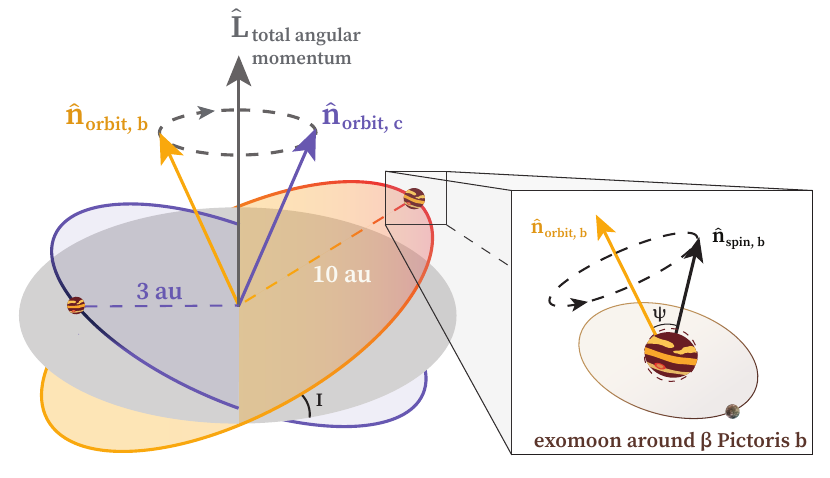}
    \caption{Schematic diagram (not to scale) of four angular momentum vectors in the $\beta$~Pictoris system: total angular momentum ($\hat{L}$), orbital angular momentum of planets b and c ($\hat{n}_{\rm orbit, b}$, $\hat{n}_{\rm orbit, c}$), and the spin vector of planet b ($\hat{n}_{\rm spin, b}$). The mutual inclination between $\hat{n}_{\rm orbit, b}$ and $\hat{n}_{\rm orbit, c}$ is $\sim 1^\circ$. Secular spin-orbit resonance can occur when the period of orbit nodal regression (path along grey dotted circle on the left) has a similar period to the spin axis precession (path along black dotted circle on the right). The period of spin axis precession can be greatly shortened in the presence of a close-in equatorial exomoon, which acts as a lever and effectively extends the planet's equatorial bulge. $I$ is the orbital inclination of $\beta$~Pictoris~b (orange ellipse) relative to the invariable plane (grey ellipse).
    }
    \label{fig:BetaPic_diagram}
\end{figure*}

The rotation period $P_{\rm rot}$ of $\beta$~Pictoris~b will yield the planet's equatorial velocity $v$, given as:
\be
    v = \frac{2\pi R}{P_{\rm rot}},
    \label{eq:equatorial_velocity}
\ee
where $R$ is the radius of $\beta$~Pictoris~b. Combining this with the independently measured \textit{line-of-sight projected} equatorial velocity $v\sin{i_p}=19.0 \pm 1.0 \units{km~s^{-1}}$ \citep{Landman+2024} from spectra will constrain the planet spin axis inclination $i_p$. Then, along with the measured orbital inclination $i_o=88.94^\circ \pm 0.02^\circ$ \citep{Brandt+2021_OrbitFits}, the line-of-sight obliquity $|i_o - i_p|$ can be calculated, which is a lower limit on the true obliquity in 3D space.

Depending on the measured rotation period, we highlight two limiting cases for the planet obliquity constraint. \autoref{fig:skyplane_obliquity} shows these two scenarios, where if $\beta$~Pictoris~b has a fast rotation period relative to its projected equatorial velocity (i.e. $v \gg v\sin{i_p}$), then the spin axis inclination $i_p \sim 0^\circ$, the projected obliquity $|i_o - i_p| \sim 90^\circ$, and its obliquity must be misaligned. Alternatively, if $\beta$~Pictoris~b spins slow relative to its projected equatorial velocity ($v \approx v\sin{i_p}$), then the spin axis inclination $i_p \sim 90^\circ$, and the \textit{line-of-sight} projected obliquity $|i_o - i_p| \sim 0^\circ$. In this case, the true obliquity in 3D space is equal to the \textit{sky-plane} projected obliquity which can range from $0^\circ-180^\circ$ and is currently not observable. This would imply that the obliquity is consistent with alignment or misalignment. In both cases, a large obliquity is possible, which motivates the following question: if $\beta$~Pictoris~b has a large obliquity, what could be the cause?

In Section \ref{sec:observations}, we explore the parameter space of possible planetary obliquity measurements given simulated rotation period measurements. 
In Section \ref{sec:obliquity_evolution}, we explore two commonly invoked explanations for Solar System planet obliquities in the context of $\beta$~Pictoris~b: collision and spin-orbit resonance.
We calculate the likelihood of collisions and then investigate if the presence of an exomoon (as in \autoref{fig:BetaPic_diagram}) can induce a secular spin-orbit resonance between the spin axis precession and orbit nodal regression of $\beta$~Pictoris~b, and therefore excite a large obliquity.
Section \ref{sec:discussion} discusses these exomoon properties in the context of planet formation and the feasibility of detection.
We summarize our conclusions in Section \ref{sec:conclusions}.

\section{Planet Obliquity Predictions} \label{sec:observations}

Preceding a concrete rotation period measurement with JWST, we calculate the planet obliquity posterior for simulated rotation period measurements.
Observations of planetary mass-companions around young stars find that objects like $\beta$~Pictoris~b typically rotate at $10-30\%$ of their break-up velocity \citep{Bryan+2018}. 
For $\beta$~Pictoris~b, \citet{Bryan+2020_AsTheWorldsTurn} use measurements of $v\sin{i_p}$, and assuming that the planet spin axis is randomly oriented in 3D space (i.e. samples of $i_p$ are drawn from a uniform distribution in $\cos{i_p}$), find that $\beta$~Pictoris~b would have an equatorial velocity corresponding to $24 \substack{+0.05 \\ -0.07} \%$ of its break-up velocity.

Motivated by these estimates, and to probe a wide range of parameter space, we generate planet obliquity posteriors for simulated rotation period measurements corresponding to 10\%, 20\%, 30\%, 50\%, or 80\% of the break-up velocity. 
We assume that this rotation period uncertainty follows a Gaussian distribution with a conservative standard deviation of 10\%. 
Early works measure the radius of $\beta$~Pictoris~b to be $1.45 \pm 0.02 \units{R_{\rm Jup}}$ \citep{Morzinski+2015}, and $1.46 \pm 0.01 \units{R_{\rm Jup}}$ \citep{Chilcote+2017}, assuming warm- or hot-start evolutionary models, but these uncertainties do not include model-dependent errors. Most recently, \citet{Kammerer+2024} use new JWST/NIRCam photometric measurements alongside three different model atmospheres, and find a wider range of possible radii spanning $\sim 1.4-1.9\units{R_{\rm Jup}}$. We choose a value in the middle of this range, $1.6\units{R_{\rm Jup}}$, and adopt a Gaussian error of $0.2\units{R_{\rm Jup}}$ to account for model-dependent errors. 
We note that this will not be the dominant source of uncertainty for the planet obliquity posterior, which comes from uncertainty of the planet spin axis orientation in the sky-plane.

\begin{figure}
    \centering
    \includegraphics[width=1.0\linewidth]{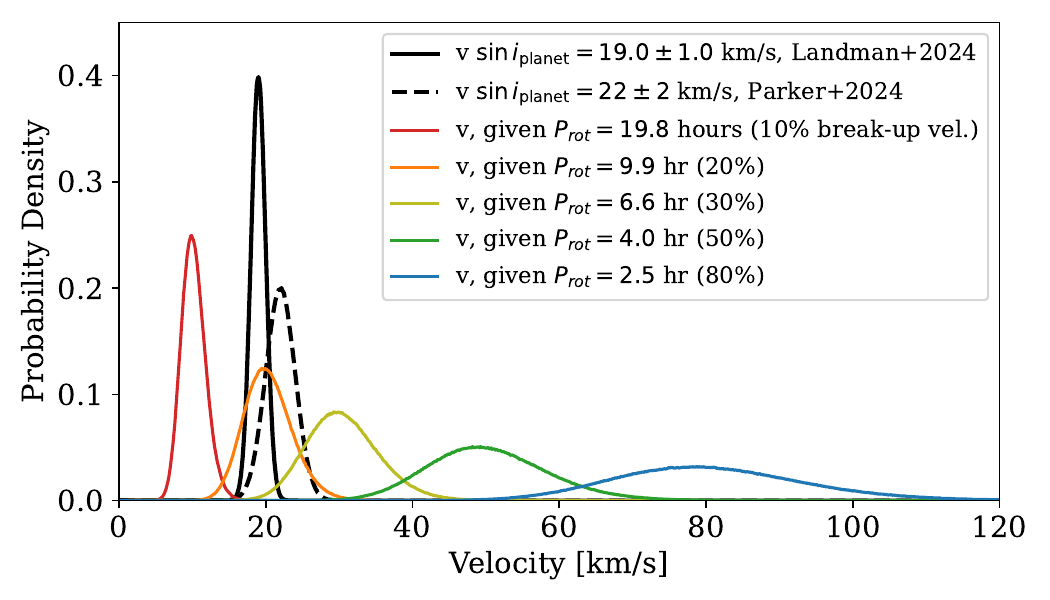}
    \caption{Posterior distributions of the measured projected equatorial velocity $v \sin{i_p}$ for $\beta$~Pictoris~b in black and black dotted. In color are deprojected equatorial velocities $v = \frac{2\pi R}{P_{\rm rot}}$, assuming $R=1.6 \pm 0.2 \units{R_{\rm Jup}}$ and simulated rotation period measurements.
    }
    \label{fig:rotation_period}
\end{figure}

Using Equation~\eqref{eq:equatorial_velocity}, we then calculate the equatorial velocity $v$ shown in color in \autoref{fig:rotation_period}.
Assuming the mass of $\beta$~Pictoris~b is $9.3 \substack{+2.6 \\ -2.5} \units{M_{\rm Jup}}$ \citep{Brandt+2021_OrbitFits}, the break-up velocity is:
\be
v_{\rm break-up} = \sqrt{\frac{GM}{R}} \approx 100 \pm 16 \units{km~s^{-1}},
\ee
where $G$ is the gravitational constant, and $M, R$ are the mass and radius of $\beta$ Pictoris b, respectively. 
This break-up velocity corresponds to a break-up rotation period of 2.0 hours.

We compare measured projected equatorial velocities \citep{Landman+2024, Parker+2024} with simulated equatorial velocities in \autoref{fig:rotation_period}. We find that a simulated rotation period corresponding to 10\% break-up velocity (red curve) would imply that the equatorial velocity is $v \approx 10 \pm 1.6 \units{km~s^{-1}}$, conflicting with the projected equatorial velocity measurement $v\sin{i_p}=19.0 \pm 1.0 \units{km~s^{-1}}$. This means that actual equatorial velocity is less than the projected equatorial velocity (i.e. $v < v\sin{i_p}$), which is unphysical. Therefore, the true rotation period likely corresponds to $\gtrsim 20\%$ of the break-up velocity.

\begin{figure}
    \centering
    \includegraphics[width=1.0\linewidth]{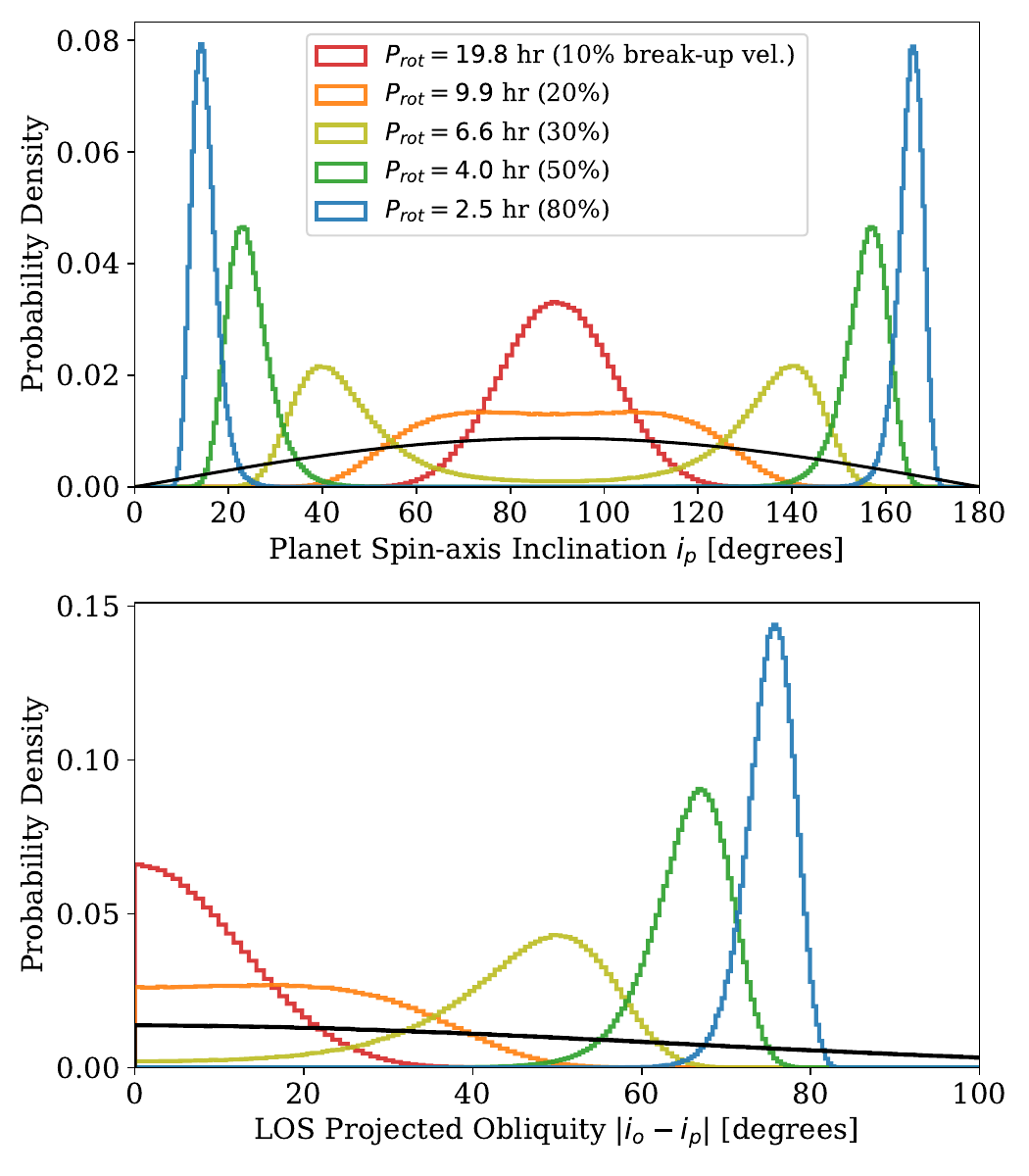}
    \caption{Top panel: Posterior distributions of spin axis inclination $i_p$ for $\beta$~Pictoris~b assuming various rotation period measurements. These distributions are compared to a random inclination distribution (black), which is uniform distribution in $\cos{i_p}$. In contrast, the orbital inclination $i_o$ is well constrained ($88.94^\circ \pm 0.02^\circ$, \citealt{Brandt+2021_OrbitFits}).
    Bottom panel: Posterior distributions for the line-of-sight projected obliquity assuming various rotation period measurements. These are lower limits on the true planet obliquity $\psi$. These distributions are compared to a randomly oriented projected inclination distribution (black), where $i_o$ and $i_p$ have both been drawn from uniform distributions in $\cos{i}$.
    }
    \label{fig:LOS_inclination}
\end{figure}

Next, we calculate the spin axis inclination in the top panel of \autoref{fig:LOS_inclination} by combining $v$ from the simulated rotation periods with $v\sin{i_p}=19.0 \pm 1.0 \units{km~s^{-1}}$ \citep{Landman+2024}. To account for correlations between $v$ and $v\sin{i_p}$, we follow the methodology of \citet{Masuda+Winn2020}. This result is also visualized in \autoref{fig:skyplane_obliquity}, which shows that zero obliquity is only possible if $i_p$ is edge-on, since $i_o$ is edge-on. However, an edge-on $i_p$ would require a slow rotation period ($\sim 10\%$ break-up velocity), which was previously discussed to be unphysical. Therefore, zero obliquity is unlikely. 

Then, we calculate the line-of-sight projected obliquity in the bottom panel of \autoref{fig:LOS_inclination}, which is the lower limit on the true obliquity in 3D space. If $\beta$~Pictoris~b spins fast ($\gtrsim 30\%$ break-up velocity), then zero obliquity is highly unlikely.

\begin{figure}
    \centering
    \includegraphics[width=1.0\linewidth]{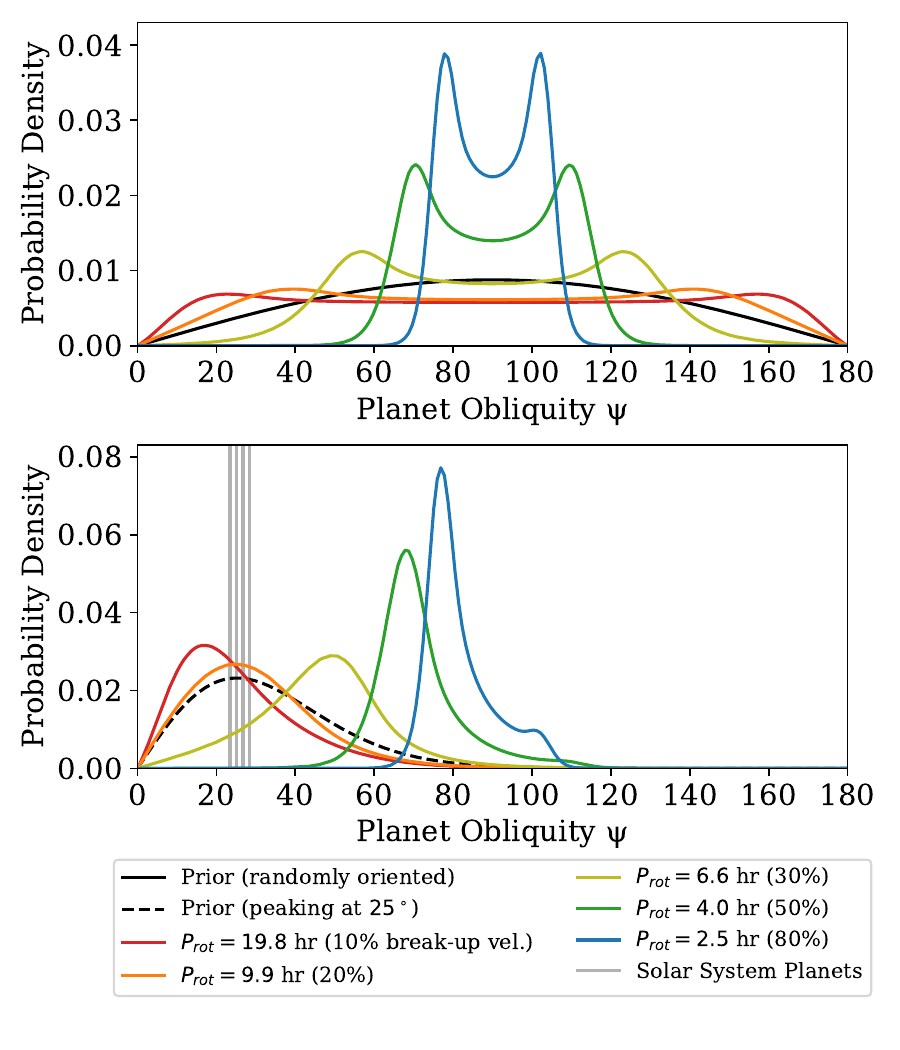}
    \caption{Top panel: Normalized posterior distributions for the planet obliquity assuming various rotation period measurements, and using a randomly oriented prior (black), which is uniform in $\cos{\psi}$. 
    Bottom panel: Same as top panel, but with a different prior (dashed black). This prior is a Fisher distribution that peaks at $25^\circ$, which is chosen to roughly match Solar System planet obliquities (from small to large: Earth, Mars, Saturn, Neptune). Mercury, Venus, Jupiter and Uranus are excluded as they have planet obliquities near $0^\circ, 90^\circ$, and $180^\circ$, preventing a simple unimodal prior.
    }
    \label{fig:obliquity_comparison}
\end{figure}

Finally, we calculate the planet obliquity in \autoref{fig:obliquity_comparison}, following the methodology of \citet{Poon+2024}, which allows for a flexible prior. We use two different priors, each highlighting a different physical theory of formation. The randomly oriented prior is motivated by top-down, star-like formation, and the Fisher distribution peaking at $25^\circ$ is motivated by bottom-up, planet-like formation. The broad uncertainty in the planet obliquity is dominated by the lack of any observable to probe the sky-plane projected obliquity. Therefore, the choice of prior can shift the peak and change the shape of the planet obliquity posterior, as shown in \autoref{fig:obliquity_comparison}.

To summarize, we do not expect $\beta$~Pictoris~b to spin very slow at $\lesssim 10\%$ break-up velocity, since the measured line-of-sight projected velocity $v\sin{i_p}$ is fast and this combination would imply an unphysical property ($v < v\sin{i_p}$). If $\beta$~Pictoris~b spins at $\sim 20\%$ break-up velocity, a small obliquity ($\sim 10-20^\circ$) is possible, but a large obliquity is much more likely. If $\beta$~Pictoris~b spins fast at $\gtrsim 30\%$ break-up velocity, only a large obliquity is possible. In all physical scenarios, a large obliquity is possible. This motivates further investigation in Section \ref{sec:obliquity_evolution} to determine what could cause obliquity evolution for $\beta$~Pictoris~b.

\section{Planet Obliquity Evolution} \label{sec:obliquity_evolution}

\subsection{Collision} \label{sec:collision}
One commonly invoked scenario to excite a large obliquity is a collision with another planet.
For example, this has been proposed as a viable scenario in the Solar System, where an Earth-mass object was proposed to have collided with Uranus during the later stages of planet formation \citep[e.g.][]{Safronov1969}.
The Safronov number $\Theta$ can be used to estimate the likelihood of collision \citep[see also][]{Tremaine2023}.
This ratio relates the escape velocity on the surface of the planet to the escape velocity at a given semimajor axis due to the star:
\be
\Theta \equiv \left(\frac{ v_\text{esc, $\beta$ Pic b} }{ v_\text{esc,$\star$} } \right)^2 = \frac{ m_2}{M_\star}\frac{a_2}{R_2} \sim 65
\ee
with $m_2$ and $M_\star$ as the planet and star mass, respectively, and $a_2$ and $R_2$ as the planet semimajor axis and radius, respectively.
We use values detailed in Table \ref{tab:beta_pic_properties}.
Since $\Theta \gg 1$, it is more likely that if an object interacts gravitationally with $\beta$~Pictoris~b, it would first have its velocity kicked $> v_\text{esc,$\star$}$ (ejection), before being kicked to $> v_\text{esc, $\beta$ Pic b}$ (collision). We also note that this $\Theta$ is greater than all of the Solar System planet $\Theta$'s (for Uranus, $\Theta \sim 5$), and we conclude that a collision at its present-day orbit is unlikely, unless the encounter is fine-tuned.

\begin{deluxetable}{ll}
\tablecaption{Measured properties of the $\beta$~Pictoris system}
\tablehead{
    \colhead{Property} & \colhead{Measurement}
    }
\startdata 
    Stellar mass ($M_\star$) & $1.83 \pm 0.04 \units{\Msun}$ \\
    Planet c mass ($m_1$) & $8.3 \pm 1.0 \units{M_{\rm Jup}}$ \\
    Planet c semimajor axis ($a_1$) & $2.74 \pm 0.03 \units{au}$ \\
    Planet c orbital period & $3.35 \pm 0.05 \units{years}$ \\
    Planet c eccentricity ($e_1$) & $0.21 \substack{+0.16 \\ -0.09}$ \\
    Planet b mass ($m_2$) & $9.3 \substack{+2.6 \\ -2.5} \units{M_{\rm Jup}}$ \\
    Planet b semimajor axis ($a_2$) & $10.26 \pm 0.10 \units{au}$ \\
    Planet b orbital period & $24.3 \pm 0.3 \units{years}$ \\
    Planet b eccentricity ($e_2$) & $0.119 \pm 0.008$ \\
\enddata
\tablecomments{From Table~1 of \citet{Brandt+2021_OrbitFits}, these measurements cite the median and 68.3\% confidence interval.}
\label{tab:beta_pic_properties}
\end{deluxetable}

\subsection{Secular spin-orbit resonance} \label{sec:resonance}

Secular spin-orbit resonance is another commonly involved mechanism to tilt a planet, and in contrast to a collision, it occurs on a timescale much longer than an orbital period. 
In our Solar System, this has been suggested to explain why Saturn has an obliquity of $27^\circ$ due to a timescale similarity between Saturn's spin-axis precession and Neptune's orbit nodal regression \citep[e.g.][]{Ward+Hamilton2004, Hamilton+Ward2004, Saillenfest+2021_Saturn, Wisdom+2022}.

Specifically, a necessary (but not sufficient) condition for a secular spin-orbit resonance to occur is that the spin precession timescale, $T_\alpha$, matches the nodal regression timescale, $T_g$.
One mechanism to regress the longitude of ascending node of the planet, $\Omega$, is due to secular interactions, described by Laplace-Lagrange theory \citep[c.f.][]{MD1999, Tremaine2023}.
In this case, the planets experience nodal regression uniformly due to each other's secular perturbations.
To leading order, the nodal regression frequency for a two-planet system is \citep{Millholland+Laughlin2019}:
\be
    \dot{\Omega} = g_{\rm LL} = -\frac{\alpha_{12}}{4} b^{(1)}_{3/2}(\alpha_{12})  \left(\frac{m_2 n_1}{M_\star + m_1} \alpha_{12} + \frac{m_1 n_2}{M_\star + m_2} \right)
    \label{eq:nodal_regression}
\ee
where $\alpha_{12} = a_1/a_2$, the Laplace coefficient is $b^{(1)}_{3/2}(\alpha_{12}) \sim 0.92$, and $n_i^2~=~GM_\star / a_i^3$ is the mean motion of planet~$i$ around the star.
Then, the nodal regression timescale is 
\be
T_g = 2\pi/|g_{\rm LL}| \simeq 30~000 \units{years},
\ee
where in the last equality we again assumed values from Table~\ref{tab:beta_pic_properties}. Here we have made the simplifying assumption that $\beta$~Pictoris~b and c are on circular orbits, since the effect of eccentricity is not significant for secular spin-orbit resonance (see \autoref{appendix:capture_into_resonance}).

In comparison, the spin axis precession timescale is
\be
T_\alpha = 2\pi/|\alpha \cos{\psi}|,
\ee
where $\psi$ is the planet obliquity and $\alpha = f_\alpha \alpha_0$ is the spin axis precession frequency.
Without a moon or circumplanetary disk, there is no enhancement, therefore $f_\alpha=1$, and $\alpha_0$ is given by \citep{Millholland+Batygin2019}:
\be
    \alpha_0 = \frac{\pi}{P_\mathrm{rot}} \frac{ k_2}{C} \frac{M_\star}{m_2}\left(\frac{R_2}{a_2}\right)^3,
    \label{eq:alpha_0}
\ee
where $P_{\rm rot}$ is the planet rotation period, $k_2$ is the planet's Love number, $M_\star$ is the host star mass, $m_2$ is the planet mass, $R_2$ is the planet radius, $C$ is the planet's normalized moment of inertia normalized by $m_2 R_2^2$, and $a_2$ is the semimajor axis. Here we choose $k_2=0.6$ to follow Jupiter's Love number as a fiducial value \citep{Lai2021}.
If $\beta$~Pictoris~b does not have a moon or disk, and has zero obliquity:
\begin{align}
T_{\alpha} = & \; 9\units{Myr} \left(\frac{P_{\rm rot}}{9.9\units{hr}}\right) \left(\frac{1.8\units{\Msun}}{M_\star}\right) \left(\frac{m_2}{9.3\units{\Mjup}}\right)  \nonumber \\
& \times \left(\frac{1.6\units{\Rjup}}{R_2}\right)^3 \left(\frac{a_2}{10.3\units{au}}\right)^3 \left(\frac{0.6}{k_2}\right) \left(\frac{C}{0.2}\right).
\end{align}

We find $T_\alpha / T_g \approx 300 \gg 1$ in this case and secular spin-orbit resonance cannot be induced. Therefore, we require a modification to either $T_\alpha$ or $T_g$ such that $T_\alpha / T_g \approx 1$ for secular spin-orbit resonance to possibly occur. We next consider if an exomoon or circumplanetary disk can provide the appropriate modification.

\subsection{The effect of an exomoon}

In the case where there is a satellite or a circumplanetary disk, $\alpha_0$ is enhanced by \citep{Millholland+Batygin2019}:
\be
    f_\alpha = \frac{C}{C + l} \cdot \frac{k_2 f_\omega^2 + 3q}{k_2 f_\omega^2}.
\ee
Here, $f_\omega$ is the planet spin rate as a fraction of the break-up rotation rate, $l$ is the normalized angular momentum for a satellite/disk with mass $m_s$, and semimajor axis $a_s$, and $q$ is the effective quadrupole coefficient of the satellite/disk. For a single equatorial satellite, $l$ and $q$ are given by \citep{Millholland+Batygin2019}:
\begin{align}
    l &= \frac{1}{f_\omega}\left(\frac{m_s}{m_2}\right)\left(\frac{a_2}{R_2}\right)^{1/2} \nonumber \\
    q &= \frac{1} {2}\left(\frac{m_s}{m_2}\right)\left(\frac{a_2}{R_2}\right)^2.
\end{align}

Recently, $\beta$~Pictoris~b completed a Hill sphere transit, and \citealt{Kenworthy+2021} placed upper limits on the mass of circumplanetary micron-sized dust in the Hill sphere as $1.8 \times 10^{22} \units{g}$, which is a factor of $4 \times 10^{3}$ less massive than Earth's moon. Therefore, we ignore the effects of a circumplanetary disk, and shift our focus the effect of an exomoon acting as the satellite. Physically, an equatorial exomoon acts as a lever to enhance the effect of planetary oblateness (shown in \autoref{fig:BetaPic_diagram}), thereby decreasing the spin axis precession period $T_\alpha$. Although multiple exomoons would additionally decrease $T_\alpha$, it would also greatly increase the complexity when comparing theoretical predictions to N-body simulations. Therefore, we focus on the effect of a single exomoon.

Next, we consider the simple scenario where $\beta$~Pictoris~b is initialized in a state with constant $T_\alpha / T_g$ and initially zero obliquity. In order for $T_\alpha \approx T_g$ in the case of near-zero obliquity, we require a moon with $f_\alpha = |g /\alpha_0| \approx 300$. The advantage of this analysis is that it does not require fine-tuning of resonance capture. We discuss this condition for how $T_\alpha / T_g$ needs to evolve in order to reach a resonant state in \autoref{appendix:capture_into_resonance}.

All that is required in addition to the observed system parameters is the presence of an exomoon. Due to the young age of the $\beta$~Pictoris system ($23 \pm 3 \units{Myr}$, \citealt{Mamajek+2014}), we ignore any exomoon migration, with further discussion in \autoref{appendix:capture_into_resonance}. By changing the mass and semimajor axis of an exomoon, we effectively change the ratio $T_\alpha / T_g$, and explore how this affects the obliquity evolution.

The equations of motion governing the spin axis evolution due to spin axis precession and nodal regression are \citep{Ward+Hamilton2004, Lu+Laughlin2022}:
\be
    \dv{\uvec{s}}{t} = \alpha \sxn \sdotn + g\sxk,
    \label{eq:dsdt_equatorial}
\ee
where $\uvec{s}$ is the planet spin axis unit vector, $\uvec{n}$ is the planet orbit normal, and $\uvec{k}$ points in the direction of the invariable plane. $\uvec{s}$ evolves in a frame rotating with angular frequency $g=g_{\rm LL}$ about $\uvec{k}$, so that $\uvec{n}$ appears fixed.

\begin{figure}
    \centering
    \includegraphics[width=1.0\linewidth]{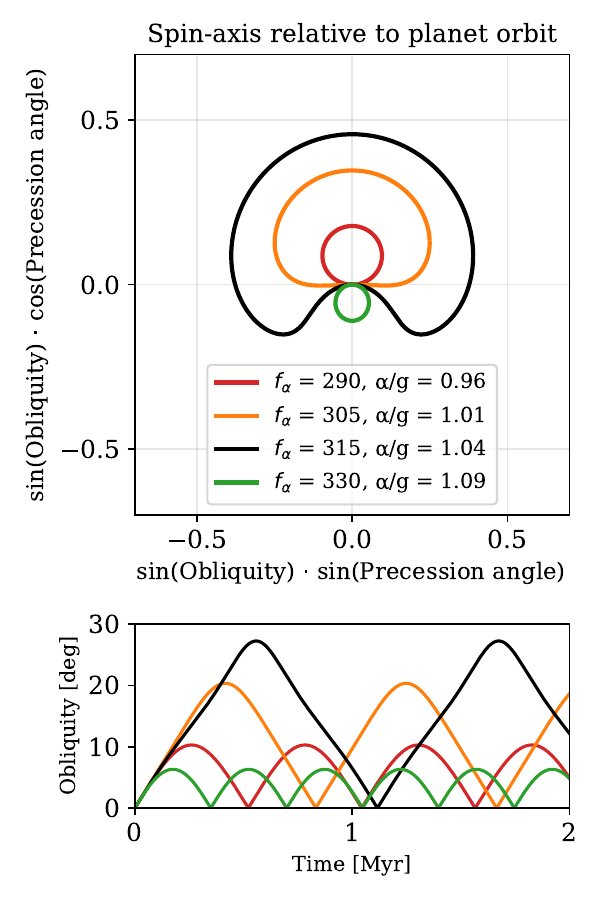}
    \caption{Top panel: Phase portrait of planet spin-axis libration (i.e. obliquity evolution) where the planet orbit normal is centred at $(0,0)$. The precession angle is the angle between the projections of the planet spin axis and invariable plane normal onto the plane of the planet orbit. The red, orange and black curves librate around Cassini state 2, whereas the green curve librates around Cassini state 1. The black curve lies near the separatrix. Bottom panel: Corresponding obliquity evolution as a function of time, initialized at zero obliquity.
    }
    \label{fig:simple_model}
\end{figure}

In \autoref{fig:simple_model}, we plot the evolution of Equation~\eqref{eq:dsdt_equatorial} initialized with various enhancement factors $f_\alpha$ such that $T_\alpha / T_g \sim 1$. We set the orbital inclination of $\beta$~Pictoris~b relative to the invariable plane $I=0.27^\circ$ (as seen in \autoref{fig:BetaPic_diagram}), calculated using the median 
measurements from Table~1 of \citet{Brandt+2021_OrbitFits}. The maximum obliquity evolution occurs near the separatrix, corresponding to \citep{Ward+Hamilton2004, Fabrycky+2007}:
\be
|\alpha / g |_{\rm crit} = \left(\sin^{2/3}{I} + \cos^{2/3}{I} \right)^{3/2} \sim 1.04,
\ee
which occurs for an equatorial exomoon with $f_\alpha=315$. Interior of the separatrix (i.e. $f_\alpha \lesssim 315$, $\alpha / g \lesssim 1.04$), the planet spin axis librates around Cassini state 2. Exterior of the separatrix (i.e. $f_\alpha \gtrsim 315$, $\alpha / g \gtrsim 1.04$), the planet spin axis librates around Cassini state 1. If $f_\alpha \ll 315$ or $f_\alpha \gg 315$, the system is not in resonance and therefore the amplitude of obliquity libration is small ($< 1^\circ$).

\begin{figure*}
    \centering
    \includegraphics[width=1.0\linewidth]{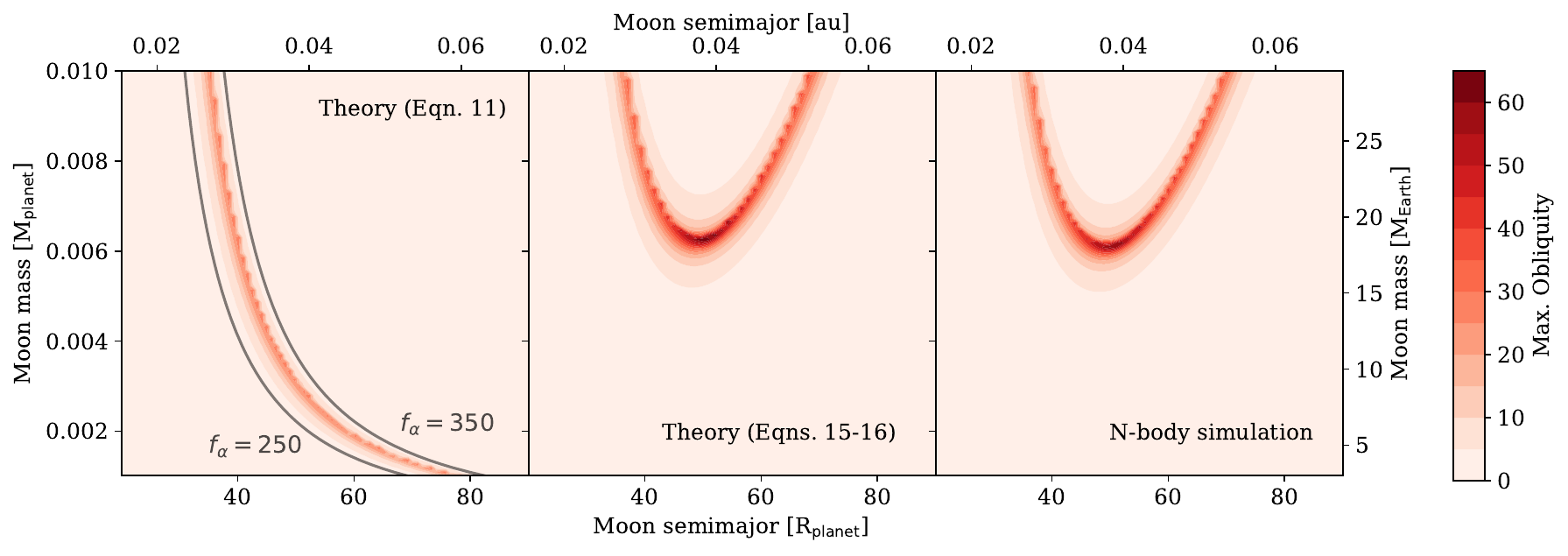}
    \caption{Maximum obliquity over $1 \units{Myr}$ for various exomoon parameters. Left: Following spin-axis equations of motion for an equatorial exomoon. Middle: Following more general spin-axis equations of motion for a non-equatorial exomoon. Here, the planet spin axis evolution is decoupled with the exomoon orbit normal evolution. Right: Comparison with N-body simulations using REBOUND, and tracking spin-axis evolution using REBOUNDx. N-body simulations (right) show strong agreement with predictions from the general spin-axis equations of motion (middle).
    }
    \label{fig:theory_vs_sims}
\end{figure*}

On the left panel of \autoref{fig:theory_vs_sims}, we plot the maximum obliquity that could be induced for a wide range of exomoon parameters, following Equation~\eqref{eq:dsdt_equatorial}. Here, we note that a less massive exomoon at a larger semimajor axis can provide the equivalent $f_\alpha$ enhancement as a more massive exomoon at a smaller semimajor axis. 

However, beyond a certain semimajor axis, an exomoon orbit normal will prefer alignment with the planet orbit normal instead of planet spin axis. This transition is known as the Laplace radius, which for a satellite on a circular orbit is (Eqn. 5.74 of \citealt{Tremaine2023}):
\begin{align}
r_L =~ &40 \units{R_2} \left(\frac{J_2}{0.008}\right)^{1/5}
                       \left(\frac{R_2}{1.6\units{\Rjup}}\right)^{2/5} \left(\frac{a_2}{10.3\units{au}}\right)^{3/5}  \nonumber \\
                       &\times  
                       \left(\frac{m_2}{9.3\units{\Mjup}}\right)^{1/5} 
                       \left(\frac{1.8\units{\Msun}}{M_\star}\right)^{1/5} 
\end{align}
where $J_2$ parameterizes the planet's oblateness, and is related to $k_2$ by (c.f. Eqn. 3 of \citealt{Millholland+Batygin2019}):
\be
J_2 = 0.008 \left(\frac{f_\omega}{0.2}\right)^2 \left(\frac{k_2}{0.6}\right).
\ee
Therefore, the left panel of \autoref{fig:theory_vs_sims} is only valid when the exomoon is interior to the Laplace radius ($a_s < r_L$).

Exterior to the Laplace radius, the exomoon orbit normal will become weakly coupled with the planet spin axis.
In this case, the equations of motion for the planet spin axis unit vector $\uvec{s}$ and the exomoon orbit normal unit vector $\uvec{l}$ are given by \citep{Tremaine1991}:
\begin{align}
    \dv{\uvec{s}}{t} =~ &\beta (h/H) \sxl \sdotl ~+  \nonumber\\
    &\alpha \sxn \sdotn + g\sxk \label{eq:dsdt}\\
    \dv{\uvec{l}}{t} =~ &\beta \lxs \lDots ~+ \nonumber\\
    &\gamma \lxn \ldotn + g\lxk. 
    \label{eq:dldt}
\end{align}
In these equations, $h=M_s M_p \sqrt{G a_s/(M_s + M_p)}$ is the reduced mass orbital angular momentum of the planet-moon system, $M_s$ is the satellite mass, and $H= 2 \pi C M_p R_p^2 / P_{\rm rot}$ is the planet spin angular momentum.
In this case, there are three precession rates relevant to the various torques in this star-planet-moon system.
Specifically, $\alpha$ is the precession rate characterizing the strength of the torque from the star on the oblate planet.
Here $\alpha=\alpha_0$ as defined previously in Equation~\eqref{eq:alpha_0}.
$\beta$ is the precession rate from the torque of the oblate planet on the satellite orbit:
\be
\beta = \frac{3}{2}n_s J_2 \left(\frac{R_p}{a_s}\right)^2,
\ee
where $n_s$ is the mean motion of the satellite around the planet.
$\gamma$ is the precession rate from the torque of the star on the satellite:
\be
\gamma = \frac{3}{4} \cdot \frac{n_p^2}{n_s}, 
\ee
where $n_p$ is the mean motion of the planet around the star.

Compared to \cite{Tremaine1991}, our equations are slightly modified by having additional $g (\uvec{s} \times \uvec{k})$ and $g (\uvec{l} \times \uvec{k})$ terms.
This is similar to how \cite{Ward+Hamilton2004} add the additional $g(\uvec{s} \times \uvec{k})$ term to Eqn.~\ref{eq:dsdt_equatorial}.
These $g$ terms account for the additional force in the rotating frame which rotates with a secular nodal regression frequency, $g = \dot{\Omega}$.
This is analogous to having additional forces when boosting to a non-inertial rotating frame.
In addition, we make a further simplification that the planet and satellite experience the same rate of nodal regression, $g$.
Hence, $g$ is the same in both differential equations.
This is a good approximation because the satellite is bound to the planet, and thus, oscillates in $\Omega$ (relative to the invariable plane) at roughly the same rate as the planet.

As seen from the set of differential equations, the planet spin axis and its exomoon orbit normal are coupled as they evolve.
In the $\gamma \ll \beta$ regime, the coupled differential equations can be shown to simplify to Equation \ref{eq:dsdt_equatorial} by observing that both $\uvec{s}$ and $\uvec{l}$ precess as a unit \citep{Tremaine1991}.
Physically, the planet spin axis and exomoon orbit normal are strongly coupled and any satellite following this criterion is usually referred to as an ``inner'' satellite.
When the criterion fails, a satellite orbits beyond the Laplace radius and is called an ``outer'' satellite.
In this regime, the full set of coupled differential equations must be solved to accurately describe evolution of the planet spin axis and exomoon orbit normal.

On the middle panel of \autoref{fig:theory_vs_sims}, we plot the maximum obliquity using Equations~\eqref{eq:dsdt}~and~\eqref{eq:dldt}, which have an overall qualitatively different behaviour than the left panel, which uses Equation ~\eqref{eq:dsdt_equatorial}.
Beyond the Laplace radius, the exomoon orbit normal becomes weakly coupled with the planet spin axis, and therefore more exomoon mass is required to induce the same planet obliquity.

\subsection{N-body comparison} \label{sec:nbody}

We verify our theoretical predictions for the exomoon-induced obliquity evolution by running direct N-body simulations with REBOUND \citep{REBOUND}, using the IAS15 integrator \citep{IAS15, Pham+2024}. We track the spin evolution of $\beta$~Pictoris~b due to rotational flattening using REBOUNDx \citep{REBOUNDx, Lu+2023}, set with a Love number of $k_2 = 0.6$ and a rotation rate at 20\% break-up velocity. We treat $\beta$~Pictoris, $\beta$~Pictoris~c, and the exomoon around $\beta$~Pictoris~b as point particles. We initialize the N-body simulation with the median parameters from Table~1 of \citet{Brandt+2021_OrbitFits} (summarized in our Table \ref{tab:beta_pic_properties}), with the modification that both planetary orbits are circular, since the effect of eccentricity is not significant for spin-orbit resonance (see \autoref{appendix:capture_into_resonance}). The exomoon is initialized on a circular orbit having an orbit normal aligned with the planet spin axis.

We run an ensemble of simulations to explore the moon semimajor and moon mass parameter space, integrating each simulation for $1 \units{Myr}$ ($\sim 30$ times the nodal regression timescale), and plot the maximum obliquity on the right panel of \autoref{fig:theory_vs_sims}. This shows excellent agreement with the middle panel of \autoref{fig:theory_vs_sims}, which resulted from evolving Equations~\eqref{eq:dsdt}~and~\eqref{eq:dldt}.

Integrating the system for longer timescales (i.e. $> 1 \units{Myr}$, as the age of system is $23 \pm 3 \units{Myr}$, \citealt{Mamajek+2014}) does not yield qualitatively different results except for the exomoon parameter space near $\sim 18 \units{M_{\rm Earth}}$ and $\sim 50$ planet radius. In some cases, the obliquity evolution reaches $\sim 90^\circ$ (polar orbit), and then appears to evolve in a complex manner. 

We additionally ran N-body simulations with no exomoon, and instead modified $k_2$ to $k_2' = f_\alpha \times k_2$, which produced results matching the left panel of \autoref{fig:theory_vs_sims}. This implies that the effect of a close-in equatorial exomoon can be reproduced by simply flattening the planet.

Our main finding is that the presence of an exomoon with mass $\gtrsim 15 \units{M_{\rm Earth}}$ ($\sim$ Neptune-mass) on an initially circular and equatorial orbit with semimajor axis $\sim 40 - 70$ planet radii can readily excite the planet obliquity of $\beta$~Pictoris~b up to $\sim 60^\circ$ in $1 \units{Myr}$. Next, we will discuss these exomoon parameters in context of planet formation, and the potential feasibility of observing such an exomoon.

\section{Discussion} \label{sec:discussion}

We have shown that the addition of a massive exomoon in the $\beta$ Pictoris system can excite the planet obliquity of $\beta$~Pictoris~b significantly on a $\units{Myr}$-timescale. This implies that if an exomoon brings a planet into a secular spin-orbit resonance, then planet obliquities can be used to indirectly constrain properties of a potential exomoon. A viable exomoon for $\beta$~Pictoris~b would need to reside within a narrow region of moon mass and semimajor axis parameter space, as shown in \autoref{fig:theory_vs_sims}.

Although there are no confirmed exomoons to date, each of the giant planets in our Solar System have more than a dozen moons. All the known moons in our Solar System have moon mass to planet mass ratio of $\lesssim 2 \times 10^{-4}$, with the exception of Earth's Moon, with ratio $\sim 10^{-2}$.  For $\beta$~Pictoris~b, we find that the exomoon to planet mass ratio required is $\gtrsim 6 \times 10^{-3}$, implying a Neptune-mass exomoon. The viable range of moon semimajor axis is $\sim 40 - 70$ planet radius (see \autoref{fig:theory_vs_sims}), which is at or beyond the Laplace radius. In contrast, most of the major moons in Solar System lie within the Laplace radius, with Earth's moon and Iapetus as the exceptions.

Recently, there have been a few exomoon candidates proposed (e.g. Kepler-1625 b-i, \citealt{Teachey+Kipping2018} and Kepler-1708 b-i, \citealt{Kipping+2022}), many of which may have a large mass.
For example, \cite{Teachey+Kipping2018} proposed an exomoon candidate Kepler-1625 b-i which has a mass of $\sim 15\units{M_\mathrm{Earth}}$ and radius of $\sim 4\units{R_\mathrm{Earth}}$, comparable to Neptune or Uranus.
Surprisingly, this candidate exomoon orbits a Jupiter-radius planet, with an upper limit of $\sim 11.6\units{M_\mathrm{Jup}}$ from radial velocity observations \citep{Timmermann+2020}. Assuming Kepler-1625 b has a mass between $1- 11.6\units{M_\mathrm{Jup}}$, this corresponds to moon to planet mass ratio of $0.4-5\%$.

First, this shows that our considered parameter space of moon masses between $\sim 1-30\units{M_\mathrm{Earth}}$ is not unprecedented.
Second, as mentioned by \cite{Teachey+Kipping2018}, this is a potential problem for existing moon formation scenarios.
\citet{Tokadjian+Piro2022} describe three possible pathways for moon formation: in situ disk accretion (e.g. theorized for the Galilean moons of Jupiter), direct impact (e.g. theorized for the Earth-Moon system), and satellite capture (e.g. theorized for Neptune's most massive and highly inclined moon Triton). 
Although in situ disk accretion is thought to create relatively low-mass moons \citep{Canup+Ward2006}, \citealt{Moraes+VieraNeta2020} argue that the presence of a massive protoplanetary disk could imply a massive circumplanetary disk, although exact properties and scaling relations are uncertain.
From numerical simulations through a population synthesis of the Galilean moons, a 1.5\% mass ratio is likely the very upper range of possible in-situ moon mass \citep{Cilibrasi2018}.
Impacts and captures are other possible ways to obtain a massive moon.
However, these scenarios need to be carefully analyzed on a system-by-system basis, since they are stochastic events.
Furthermore, it is unclear on how common impacts or captures are in early planetary systems.
Each of these scenarios will have different implications on the tilts of exoplanets and satellites \citep{Teachey+Kipping2018}.
Alternatively, the orbital architecture of planets around late M-dwarfs like the TRAPPIST-1 system (7 Earth-sized planets around a $\sim 80$ Jupiter mass M-dwarf at $\sim 20-100$ stellar radius) can be a useful analog in understanding the formation of large exomoons \citep{Ormel+2017}.

Finally, we discuss the feasibility of detecting a Neptune-mass exomoon around $\beta$~Pictoris~b. 
Assuming a uniform density of $1-3 \units{g~cm^{-3}}$, which accounts for a range of Neptune-like to rock-like densities, a Neptune-mass moon would have a radius of $\sim 3.2-4.6$ Earth radius. 
If the moon transits in front of a $1.6$ Jupiter radius planet, it would have a transit depth of $3-7\%$. 
Such a transit depth would likely dominate the variability amplitude (typically $\sim 0.2\% - 2\%$, \citealt{Metchev+2015}) of a planetary-mass object or a variable brown dwarf due to rotation, and is possibly detectable with JWST \citep{Cassese+2022, Liu+2024}.
With a semimajor axis of $\sim 40 - 70$ planet radius, this exomoon would have an orbital period of $3-7$ weeks. 
However, the chance of transit will be low if $\beta$~Pictoris~b has a nonzero obliquity, since our proposed mechanism implies that the exomoon orbital plane is likely inclined relative to the edge-on planetary orbit.

\section{Conclusions} \label{sec:conclusions}

In this study, we predict the planet obliquity posterior of $\beta$~Pictoris~b, show that the presence of an exomoon can explain a nonzero planet obliquity, and explore the exomoon parameter space. 
Our key findings are as follows:

1. If $\beta$~Pictoris~b spins fast ($\gg 20\%$ break-up velocity, or $P_{\rm rot} \ll$ 9.9 hours\footnote{
    This rotation period calculation assumes a radius of $1.6 \units{\Rjup}$. Using $1.5 \units{\Rjup}$ instead decreases this rotation period to 8.9 hours.}, the planet obliquity must be misaligned. If $\beta$~Pictoris~b spins slow ($\lesssim 20\%$ break-up velocity, or $P_{\rm rot} \gtrsim$ 9.9 hours) the planet obliquity can be aligned or misaligned. Further constraints would need to probe the sky-plane projected obliquity.

2. The presence of an exomoon can cause a secular spin-orbit resonance for $\beta$~Pictoris~b, and excite its planet obliquity from $0^\circ$ up to $\sim 60^\circ$ within $1 \units{Myr}$.
We use a general framework that allows the exomoon to be close-in (equatorial) or distant (not equatorial), and find that our analytic predictions match with our N-body simulations.

3. This exomoon needs to be have at least one Neptune-mass, and have a semimajor axis of $\sim 40 - 70$ planet radius ($\sim 0.03 - 0.05 \units{au}$).

The secular spin-orbit resonance mechanism we present can be applied to a broad range of multi-planet systems, where the presence of an exomoon can excite a planet obliquity significantly. 
A nonzero planet obliquity constraint can potentially provide concrete predictions for viable exomoon parameters, and provide clues about the extremes of moon and planet formation beyond our Solar System.

\begin{acknowledgments}
We thank the anonymous reviewer for their helpful comments which have greatly improved the clarity of our manuscript.
We thank Marta Bryan, Mark Dodici, Tiger Lu, Sarah Millholland, Yubo Su, and Scott Tremaine and  for useful discussions.
\end{acknowledgments}

\appendix

\section{Capture into secular spin-orbit resonance} \label{appendix:capture_into_resonance}

A secular spin-orbit resonance occurs when the ratio of the spin axis precession timescale to the nodal regression timescale $T_\alpha / T_g$ evolves through unity from above, and that this resonance-crossing timescale slower than the resonant liberation timescale (Eqn. 8 of \citealt{Hamilton+Ward2004}):
\be
T_{\rm lib} = 0.5\units{Myr}
\left(\frac{2\pi/\alpha}{30~000 \units{yr}}\right)^{1/2}
\left(\frac{2\pi/|g|}{30~000 \units{yr}}\right)^{1/2}
\left(\frac{10^\circ}{\psi}\right)^{1/2}
\left(\frac{1^\circ}{I}\right)^{1/2}
\ee

In the following, we investigate how the eccentricity of $\beta$~Pictoris~b and $\beta$~Pictoris~c, as well as the possibility of planet or moon migration, can aid capture into secular spin-orbit resonance, and ultimately affect the planet obliquity of $\beta$~Pictoris~b.

\subsection{The effect of eccentricity} \label{appendix:eccentricity}

In all the simulations of this work, we assume that $\beta$~Pictoris~b \& c have circular orbits. 
Although this simplifies the dynamics, both these planets have nonzero eccentricities. \citet{Brandt+2021_OrbitFits} find an eccentricity of $0.119 \pm 0.008$ for $\beta$~Pictoris~b and $0.21 \substack{+0.16 \\ -0.09}$ for $\beta$~Pictoris~c. 
Including the effects of eccentricity, $g_{\rm LL}$ from Equation~\eqref{eq:nodal_regression} is modified by a factor of $(1-e_1^2)^{-1/2}$ in the first term and $(1-e_2^2)^{-1/2}$ in the second term (e.g. Equations~4~and~5 in \citealt{Bailey+Fabrycky2020}, \citealt{MD1999}). 
Similarly, $\alpha_0$ from Equation~\eqref{eq:alpha_0} is modified by a factor of $(1-e_2^2)^{-3/2}$ (e.g. Equation~12 in \citealt{Bryan+2020_Obliquity}). 
Although these modifications change $T_\alpha / T_g$ by a few \%, the eccentricity evolves on a timescale $T_g \ll T_{\rm lib}$, and $T_\alpha / T_g$ does not evolve monotonically, but rather oscillates. 
Therefore, the effect of eccentricity is not be significant to the planet obliquity evolution of $\beta$~Pictoris~b.

\subsection{The effect of planet migration} \label{appendix:planet_migration}

\begin{figure}
    \centering
    \includegraphics[width=1.0\linewidth]{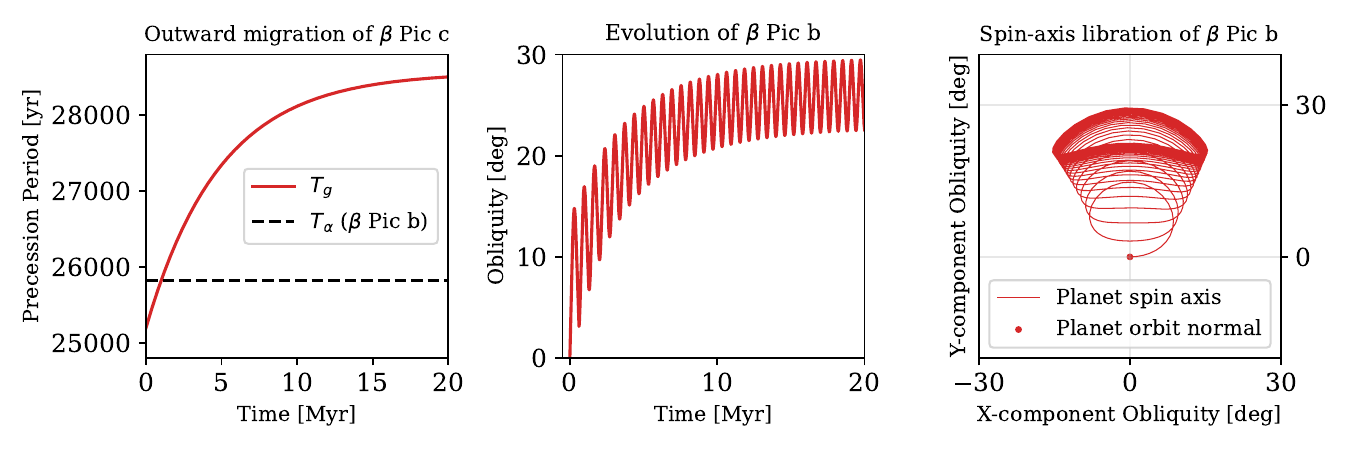}
    \caption{Left: As $\beta$~Pictoris~c migrates outwards from $2.4 \units{au}$ to present-day $2.7 \units{au}$ over a timescale of $5 \units{Myr}$, the nodal regression timescale $T_g$ increases. Middle: Simultaneously, $\beta$~Pictoris~b becomes captured into a secular spin-orbit resonance, and its planet obliquity is excited to $\sim 25^\circ$. Right: During this capture, the spin axis of $\beta$~Pictoris~b evolves from initial alignment with the orbit normal, to libration around Cassini state 2 in a banana-like trajectory.
    }
    \label{fig:migration}
\end{figure}

In the presence of an exomoon, one way to satisfy the capture criterion is to have $\beta$~Pictoris~c initially interior to its present day orbit, and migrate outwards. 
Then, $T_g$ would decrease during migration while $T_\alpha$ remains constant. 
In \autoref{fig:migration}, we show that an exomoon contributing a spin axis precession enhancement of $f_\alpha=340$ in addition with a simple exponential migration model from $2.4 \units{au}$ to $2.7 \units{au}$ over a timescale of $5 \units{Myr} > T_{\rm lib}$, can provide access to large planet obliquities for $\beta$~Pictoris~b.

\subsection{The effect of moon migration} \label{appendix:moon_migration}

Another way to satisfy resonance-crossing is to instead have an exomoon around $\beta$~Pictoris~b migrate outwards, thereby decreasing $T_\alpha$ while $T_g$ remains constant. 
This mechanism has been proposed to explain Saturn's large planet obliquity, with the migration of Titan \citep{Saillenfest+2021_Saturn, Saillenfest+2021_Saturn2, Wisdom+2022}, as well as suggest that Jupiter's planet obliquity may increase as the Galilean moons migrate outwards. 
However, $\beta$~Pictoris~b is much more massive than the Solar System gas giants, making it harder to tilt, and this mechanism would have much less time to take effect, as the $\beta$~Pictoris system is relatively young ($23 \pm 3 \units{Myr}$, \citealt{Mamajek+2014}). 
Therefore, we expect that this mechanism is not significant, unless moon migration is exceptionally fast.

\vspace{10cm}

\bibliography{main}{}
\bibliographystyle{aasjournal}

\end{document}